\documentclass[a4paper]{toledoStyle} 
\usepackage{epsfig}

\begin{document}

\title{The far infrared CO line emission of Orion BN/KL}

\author{S. Maret\inst{1} \and E. Caux\inst{1} \and J.P. Baluteau\inst{2} \and
  C. Ceccarelli\inst{3} \and C. Gry\inst{2,4} \and C. Vastel\inst{1}}

\institute{CESR CNRS-UPS, BP 4346, F-31028 - Toulouse cedex 04, France \and
  Laboratoire d'Astrophysique de Marseille, Traverse du siphon, F-13004
  Marseille, France \and Laboratoire d'Astrophysique, Observatoire de Grenoble
  -BP 53, F-38041 Grenoble cedex 09, France \and ISO Data Center, ESA
  Astrophysics Division, Villafranca del Castillo, P.O. Box 59727 Madrid,
  Spain }

\maketitle

\begin{abstract}
  We present observations of the closest region of high mass star formation,
  Orion BN/KL, performed at both low resolution mode (grating mode) and high
  resolution (Fabry-P\'erot mode) with the Long Wavelength Spectrometer (LWS)
  on board the Infrared Space Observatory (ISO). We detected the CO rotational
  transitions from J$_{up}$= 15 to J$_{up}$ = 49.  A LVG analysis of the line
  fluxes allows to distinguish three main physical components with different
  temperatures, densities and column densities. Our analysis yields to
  conclude than J$_{up}<$15 arise from the photodissociation region (PDR), the
  emission between J$_{up}$= 20 and 30 arise from the high velocity outflow
  (plateau), and J$_{up}$ $>$ 32 arises from a hot and dense gas component.
  The latter exhibits broadened lines for the levels J$_{up}$ $>$ 32 and is
  thought to be due to shocked gas in a high velocity outflow. Future
  observations with HIFI, onboard the Far Infrared Space Telescope (FIRST)
  will allow the spectral separation of the PDR and the plateau components,
  unresolved with ISO, and characterise more precisely the Orion BN/KL star
  forming region.
\end{abstract}

\section{Introduction}
The Orion Molecular cloud, at a distance of 450 pc, is the closest region of
high mass star formation. Its proximity and its large infrared luminosity
allowed to perform plenty of observations in the past years, yielding the
discovery of the first proto-stars candidates. Molecular emission mostly comes
from OMC1, containing several condensation as the KL nebulae. The KL nebulae
is composed by many infrared clusters (e.g. BN or IrC2), containing massives
stars at early evolutionary states. Millimeter, sub-millimeter and infrared
spectroscopy (see \cite{GenzelStutzki} for a review) have shown copious
molecular emission arising from physically distinct regions: the \emph{ridge},
the \emph{compact ridge}, the \emph{hot core}, the PDR region surrounding the
quiescent gas and a high velocity ($\Delta v$ = 18 km/s) bipolar outflow
originating from IrC2.  Outside the bipolar outflow is a region of very hot
(1000 to 2000 K) shock exited gas. Numerous observations have allowed to
charaterise this components in term of temperature, density and column
density. Recently \cite*{Sempere} reported of FIR CO observations towards
BN/KL with the Long Wavelength Spectrometer (hereafter LWS: \cite{Clegg}) on
board of the Infrared Space Observatory (ISO: \cite{Kessler}). Theses
observations revealed the presence of three gas components they identified as
the ridge, the high velocity outflow and a hot and dense gas component
detected at J$_{up}$ $>$ 30 due to the shocked gas in the high velocity
outflow. Notheless uncertainties remains on the data calibration, and high J
grating data are probably contaminated by adjacent lines. We here present a
large bandwith survey of the far infrared CO lines of Orion BN/KL, performed
with ISO-LWS in Fabry-Perot mode.  Particular emphasis is given to the
calibration of the FP data. Based on theses new calibrated data we interprete
the FIR lines by means of a LVG code to derive the temperature, density and
column density of the several gas component.

\section{Observations and results}

We performed a spectral survey of the Orion BN/KL using ISO-LWS both in
grating mode and FP mode. The 80$\arcsec$ beam was centred on Orion BN/KL
($\alpha_{2000}=5^{h}35^{m}14.2^{s}$, $\delta_{2000}=-5 \degr 22 \arcmin 33.6
\arcsec$). The grating spectral survey was done using LWS in with the L01 AOT.
It was calibrated using Uranus, and the absolute accuracy is estimated to be
better than 30\% (\cite{Swinyard}). Theses observations required the use of
the bright source data reduction package (\cite{Leeks}) as the brightness of
the source saturated the detectors LW2 to LW4. The grating spectrum was only
used to calibrate the FP observations. The FP spectral survey was performed
with the L03 AOT. Theses observations, covering the wavelength range 43 to 162
$\mu$m are the first performed with a single instrument in space at the same
time. The continuum level of the observation was calibrated against the
grating spectrum, after taking into account the dark, straylight and FP order
sorting. The calibration uncertainties, estimated from RMS of measured fluxes
of each transition, is between 30\% and 40\% for the weaker lines. We detected
$^{12}$CO rotational lines between J$_{up}$ = 15 and 49 in FP mode. Higher
transitions fluxes are under 10$^{-11}$ erg.s$^{-1}$.cm$^{-2}$ and are not
detected. The fluxes uncertainties, estimated from $\sigma$ measurements of
each transitions, are about 30\% for transitions between J$_{up}$ = 15 and 40
and 40\% between 40 and 45. Lines between J$_{up}$ = 45 and 49 must be
considered as upper limits.
\begin{figure}
  \begin{center}
    \epsfig{file=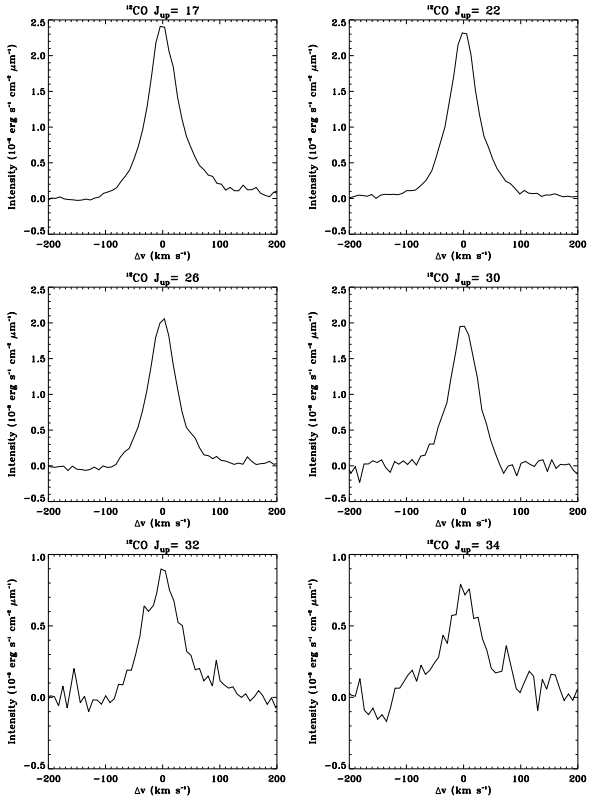, width=8cm}
  \end{center}
  \caption{Observed profiles of selected $^{12}$CO rotational transitions
    whit LWS in FP mode. J$_{up}$ $>$ 32 transitions shows an important
    broadening}
    \label{12CO lines}
\end{figure}
Higher transitions have a low RMS yielding uncertainties on width
determination but the broadening observed is consistent with higher
transitions measurement. This is clearly indicate that J$_{up}$ $\ge$ 32 lines
emission comes from another gas component than lower transitions, necessary
hotter and denser. We also detected J$_{up}$ = 18 and 24 $^{13}$CO lines. We
found a flux of respectively 3.7 $\pm$ 1.1 and 4.7 $\pm$ 1.4 erg s$^{-1}$
cm$^{-2}$. The $^{12}$CO/$^{13}$CO fluxes ratio is 27 for J$_{up}$ = 18 and 12
for J$_{up}$ = 24, indicating that theses lines are optically thick.

\section{Discussion}
\begin{figure}
\begin{center}
  \epsfig{file=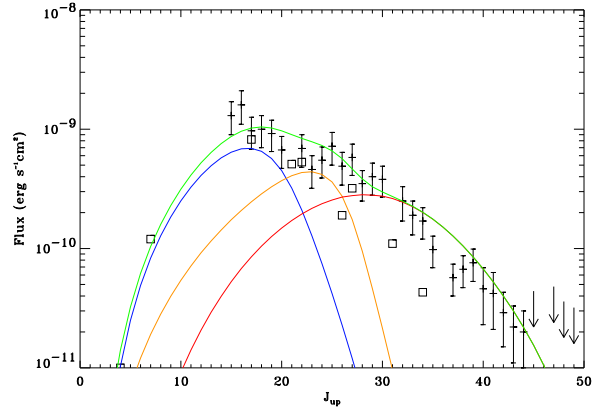, width=8cm}
\end{center}
    \caption{Observed $^{12}$CO rotational transitions by LWS in FP mode
      (crosses with error bars), and three components model of Orion
      BN/KL. Uncertainties on LWS FP lines fluxes are given at 1 $\sigma$. We
      also plotted the values from anteriors observations. The differences
      between this mesurements and LWS mesurement are probably due to the
      differents instruments beams. PDR emission is represented in blue,
      plateau emission in orange and shocked gas emission in red.  The total
      emission is represented in red. Predominance of the shocked gas
      component from J$_{up}$ = 32 allows to explain the observed broadening
      at these transitions.}
    \label{CO spectrum}
\end{figure}

We analysed the $^{12}$CO lines fluxes by means of an LVG model developed by
Ceccarelli et al. This model, which compute in a self consistent way the
opacities of lines, has four free parameters : the $^{12}$CO column density,
the H$_{2}$ density, the gas temperature and the linewidth. The results of our
computations are shown on fig. \ref{CO spectrum}, together with our
observations and previous ones (\cite{Schultz}, \cite{Graf},
\cite{Schmid-Burgk}, \cite{Howe}, \cite{Genzel}). The differences between
these previous measurements and our measurements are probably due to the
different instruments beam.
The first thing to note is that a single gas component can not explain the
J$_{up}$ = 15 to 44 observed emission.  At least two components are needed to
explain the emission peak observed at J$_{up}$ = 16 and the ``broad shaped''
emission between J$_{up}$ = 20 and 30. In addition the observed broadening at
J$_{up}$ $>$ 32 implies a third physical component.

\subsection{Low J emission. Ridge, compact ridge and PDR}
\begin{figure}
  \begin{center}
    \epsfig{file=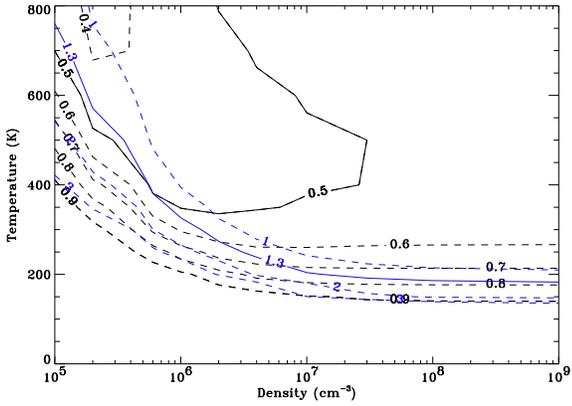, width=8cm}
  \end{center}
    \caption{Theoretical J$_{up}$ = 15 / 20 $^{12}$CO lines ratios (in blue) and photon escape probability at J$_{up}$ = 18 (in black)
      predicted by our LVG model. The observed values are represented by
      continuous lines and the errors bar by dashed lines lines. We assumed
      N(CO) = 10$^{18}$ cm$^{-2}$ and $\Delta v$ = 10 km/s. This parameters
      gives a lower limit of 250 K for the temperature and 10$^{5}$ cm$^{-3}$
      for the density. The best fit model is obtained for n(H$_{2}$) =
      10$^{6}$ cm$^{-3}$ and T = 400 K.}
    \label{PDR temperature density}
\end{figure}

Because of the accurate calibration we used, the emission between J$_{up}$ =
15 and 20 we observed is quite different than previous ISO observations. The
fluxes we measure are 30 \% higher than those measured by \cite{Sempere}. Such
fluxes can't be explain by the ridge emission as they previously said. The
temperature, density and column density they adopted fail to reproduce the
intensity of the emission we observed. In order to characterise the gas
emitting at these transitions, we used the $^{12}$CO to $^{13}$CO line ratio
at J$_{up}$ = 18 to calculate the escape photon escape probability at this
transition. Assuming that the relative abundance of $^{13}$CO with respect to
$^{12}$CO is 60 and that the $^{13}$CO is optically thin, it gives a photon
escape probability of 0.5 at J$_{up}$ = 18.  We also calculated J$_{up}$ = 15
/ 20 $^{12}$CO lines ratios. This ratio, both with the photon escape
probability were used to constrain the temperature and density of the gas. The
fig. \ref{PDR temperature density} shows theoretical photon escape probability
and lines ratios computed with our LVG codes, both with photon escape
probability and ratio we observed. We found a lower limit for the CO column
density of 10$^{17}$ cm$^{-2}$. We adopted a column density of 10$^{18}$
cm$^{-2}$ and a line width of 10 km/s. This parameters gives a a temperature
lower limit of 200 K. The best fit model is obtained for T = 350 K and
n(H$_{2}$) = 10$^{6}$ cm$^{-3}$, but observed emission can also be explain by
lower temperature and higher density, or inversely. The lack of data between
J$_{up}$ = 7 and 15 not allow to constrain more precisely the physical
parameters. The parameters we adopted require a beam dilution factor of 1,
which implies an extended emission. This, both with temperature limit of 200 K
we obtained, indicates that the emission between 15 $<$ J$_{up}$ $<$ 20 can
not only arise from the ridge, but may arise from a the PDR region, in
agreement with \cite{Howe}. Notheless, BN/KL is a complex region where a lot
of gas component whit nearby physical characteristics are present. This
components are not resolved with LWS. Even if a single gas component can
explain the J$_{up}$ = 15 to 20 emission we observed, a part of emission may
arise from the ridge. Nonetheless, because of it low temperature, we estimated
the low J emission contribution of ridge of 10 \% of the PDR emission. The hot
core, due to its small angular size towards the PDR, also certainly few
contribute to the 15 $<$ J$_{up}$ $<$ 20 emission. The density and column
density are in agreement with previous works (\cite{Howe}), but the
temperature we derive is significantly higher.

\subsection{J$ _{up}$ = 20 to 30 emission. Plateau}
Observed emission between J$_{up}$ = 20 and 30 show a broad shaped emission
necessary arising from an hotter and denser component than the PDR. Along the
same lines, we used the $^{13}$CO to $^{12}$CO fluxes at J$_{up}$ = 24 to
constrain the physical parameters of the gas emitting at theses transitions.
This ratio give a photon escape probability of 0.2. We used the $^{12}$CO line
ratio 20/26 is 1.4 and adopted a line width of $\Delta$v = 30 km/s.\\
We found a lower limit for the column density of 10$^{19}$ cm$^{-2}$, but
higher density are also giving solution. We adopted N(CO) = 5.10$^{19}$
cm$^{-2}$. This column density gives a lower limit of 5.10$^{6}$ cm$^{-3}$ for
the density and 300 K for the temperature (see fig. \ref{Plateau temperature
  density}).
\begin{figure}
  \begin{center}
    \epsfig{file=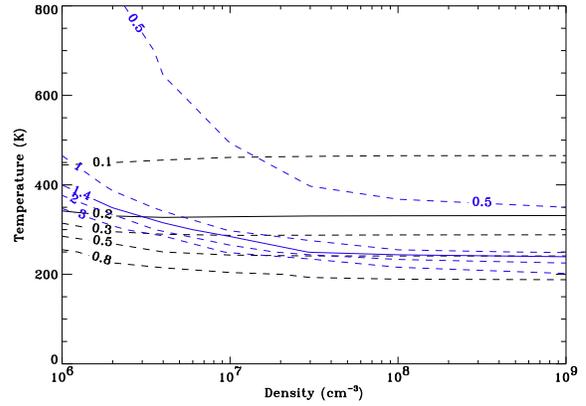, width=8cm}
  \end{center}
    \caption{Theoretical J$_{up}$ = 26 / 20 $^{12}$CO lines ratios (in blue) and
      photon escape probability at J$_{up}$ = 24 (in black) predicted by our
      LVG model. We assumed N(CO) = 5.10$^{19}$ cm$^{-2}$ and $\Delta v$ = 30
      km/s . It gives a lower limit of 5.10$^{6}$ cm$^{-3}$ for the density
      and 300 K for the temperature}
    \label{Plateau temperature density}
\end{figure}
The parameters which are best fitting our observations are T = 350 K and a
n(H$_{2}$) = 10$^{7}$ cm$^{-3}$. Theses requires a beam filling factor of 7 \%
to fit the J$_{up}$ = 20 to 30 lines fluxes. The angular size we derive from
this filling factor is 20 $\arcsec$ \cite{Howe}. The high velocity component
angular size deduced from CO 7-6 line emission is 40 $\arcsec$, but could be
smaller at higher transitions. The angular size we determined yields to say
that the emission at theses transitions arises from this gas component. The
physical parameters we derived agree with previous determinations
(\cite{Schultz1995}).

\subsection{High J emission. Shocked gas}
Although high J emission could be explain by higher plateau temperatures and
densities, the observed broadening of J$_{up}$ $>$ 32 lines clearly shows that
emission at this transitions arises from another hot and dense gas component.
High temperatures and densities that requires this emission suggest that the
emission arise from the gas shocked by the high velocity outflow, in agreement
with \cite{Sempere} interpretation.  We modelled this emission by a gas
temperature of 1500 K, n(H$_{2}$) = 4.10$^{6}$ cm$^{-3}$, and $\Delta$v = 50
km/s. Theses requires a \emph{beam averaged} column density of 10$^{17}$
cm$^{-2}$. Assuming a angular size of 40 $\arcsec$, it implies a column
density of 10$^{16}$ cm$^{-2}$. This values are in good agreement with
\cite{Sempere}.  At J$_{up}$ $>$ 32, the shocked gas emission become more
important than the plateau emission, explaining the line broadening observed
at theses transitions.

\subsection{Agreement with observed lines profiles}
\begin{figure}
  \begin{center}
    \epsfig{file=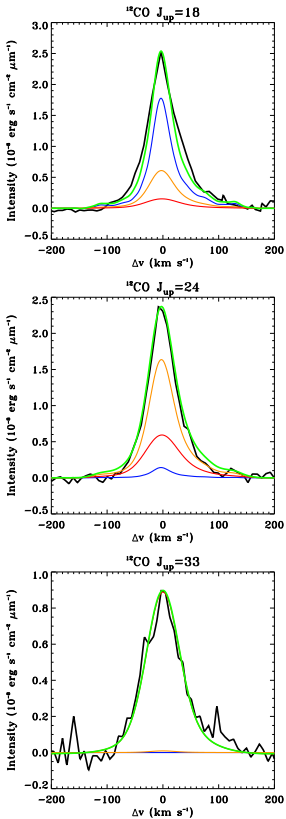, width=4cm}
    \end{center}
    \caption{J$_{up}$ = 18, 24 and 33 $^{12}$CO rotational transitions,
      superposed on PDR emission (blue), plateau (orange), shocked gas (red),
      and total (green) predicted by the LVG model. Theoretical line profiles
      have been obtained in convolving gaussians profiles of respectively 10,
      30 and 50 km/s width and of intensity deduced by the LVG model, by the
      PSF of the instrument.  Predominance of shocked gas emission at J$_{up}$
      = 32 can explain the broadening of lines observed at theses
      transitions.}
    \label{fitted 12CO lines profiles}
\end{figure}

In order to check the agreement of our model with observed lines profiles, we
compared theoretical emission lines profiles of the three component with
observed profiles. Theoretical profiles were obtained in convolving gaussian
profiles by the PSF of the instrument. The line intensities were inferred from
lines fluxes predicted by our LVG model. The PSF was obtained by Vastel et al.
in observing thin [OI],[OIII] and [CII] lines in NGC7023 and G 0.6-0.6. The
theoretical profiles, superposed on observed lines, are shown on fig.
\ref{fitted 12CO lines profiles}.
The theoretical profils are in good agreement with the observed ones. The
predominance of shocked gas emission at J$_{up}$ = 32 can explain the
broadening of lines observed at theses transitions

\section{Conclusions}
The Orion BN/KL observations with LWS allowed to detect J$_{up}$ = 15 to 49 CO
rotational transitions in Fabry-Perot mode. The modelling of observed fluxes
by a LVG showed that molecular emission can be explained by three gas
components that we characterised in term of temperature, density and column
density (see tab. \ref{Physical parameters and uncertainties}).
\begin{table}
  \begin{center}
    \caption{Temperatures,densities, column densities and micro turbulent
      velocities of ridge, PDR, plateau and the shocked gas determinated from
      our LVG model} \footnotesize
    \begin{tabular}{c c c c}
      \hline
      &PDR&Plateau&Schocked gas\\
      \hline
      n(H$_{2}$) cm$^{-3}$&10$^{6}$&2.10$^{6}$&4.10$^{6}$\\
      T (K)&400&350&1500\\
      N(CO) (cm$^{-2}$)&10$^{18}$&5.10$^{19}$&10$^{17}$\\
      $\Delta v$ (km/s)&10&30&50\\
      Filling factor (\%)&100&7&25\\
      \hline
    \end{tabular}
    \label{Physical parameters and uncertainties}
\end{center}
\end{table}
We found that low J emission can not only arise from the ridge, too cold, but
may arise from the PDR region. The emission between J$_{up}$ = 20 and 30
certainly arise from the high velocity outflow. Finally, observed J$_{up}$ $>$
32 lines broadening is though to be due to the gas shocked by the high
velocity outflow. High J fluxes measurements lead to estimate a shocked gas
temperature between 1000 and 2000 K, a H$_{2}$ density between 10$^{6}$ and
10$^{7}$ cm$^{-3}$ and a CO column density between 10$^{15}$ and 10$^{16}$
cm$^{-2}$. Our model both account of observed fluxes and lines profiles. This
study show the necessity of high resolution observations of Orion BN/KL, as
ISO-LWS observations not allow the spectral separation of the different gas
components. Future observations with HIFI, onboard the Far Infrared Space
Telescope (FIRST) will allow the spectral separation of the PDR and the
plateau component, unresolved with ISO, and characterise more precisely the
Orion BN/KL star forming region.

\begin{acknowledgements}
  This study is based on observations with ISO, an ESA project with
  instruments funded by ESA Member States (especially the PI countries:
  France, Germany, the Netherlands and the United Kingdom) with the
  participation of ISAS and NASA.
\end{acknowledgements}

\end{document}